\documentclass[11pt,a4paper]{article}
\usepackage{jheppub}
\usepackage{setspace,braket}
\usepackage{amsmath, amssymb, bm, yfonts}
\usepackage{etoolbox}

    \makeatletter
    \patchcmd{\maketitle}{\@fpheader}{}{}{}
\makeatother
\makeatletter
\newcommand{\vast}{\bBigg@{3.5}}
\newcommand{\Vast}{\bBigg@{5}}
\makeatother

\def\be{\begin{equation}}
\def\ee{\end{equation}}

\def\be{\begin{equation}}
\def\ee{\end{equation}}

\def\a{\alpha}

\def\bg{\bar{g}}
\def\beq{\begin{eqnarray}}\def\eeq{\end{eqnarray}}
\def\ba#1\ea{\begin{align}#1\end{align}}
\def\bg#1\eg{\begin{gather}#1\end{gather}}
\def\bm#1\em{\begin{multline}#1\end{multline}}
\def\bmd#1\emd{\begin{multlined}#1\end{multlined}}

\def\a{\alpha}

\def\({\left(}
\def\){\right)}
\def\[{\left[}
\def\]{\right]}

\def\Tr{{\rm Tr}}

\def\Tr{{\rm Tr }}

\def\a'{\alpha'}

\newcommand{\bea}{\begin{eqnarray}}
\newcommand{\eea}{\end{eqnarray}}

\newsavebox{\uuunit}
\sbox{\uuunit}
    {\setlength{\unitlength}{0.825em}
     \begin{picture}(0.6,0.7)
        \thinlines
        \put(0,0){\line(1,0){0.5}}
        \put(0.15,0){\line(0,1){0.7}}
        \put(0.35,0){\line(0,1){0.8}}
       \multiput(0.3,0.8)(-0.04,-0.02){10}{\rule{0.5pt}{0.5pt}}
     \end {picture}}

\numberwithin{equation}{section}
\renewcommand{\thesection}{\arabic{section}}

\begin{document}
\renewcommand{\thefootnote}{\fnsymbol{footnote}}

\title{\begin{flushright}\small  WITS-MITP-003 \end{flushright}
Free-kick condition for entanglement entropy in higher curvature gravity}
\author[a,b]{Seyed Morteza Hosseini}
\author[c]{and \'{A}lvaro V\'{e}liz-Osorio}
\affiliation[a]{Dipartimento di Fisica, Universit\`a di Milano-Bicocca,\\I-20126 Milano, Italy}
\affiliation[b]{INFN, sezione di Milano-Bicocca, I-20126 Milano, Italy}
\affiliation[c]{Mandelstam Institute for Theoretical Physics, School of Physics\\University of the Witwatersrand, WITS 2050, Johannesburg, South Africa}
\emailAdd{morteza.hosseini@mib.infn.it}
\emailAdd{alvaro.velizosorio@wits.ac.za}
\abstract{
In order to compute the entanglement entropy for a given region in a theory with an Einstein gravity dual,
the Ryu-Takayanagi prescription tells us that we must compute the area of an extremal surface {\it anchored} to the entangling region.
However, if the dual gravity theory receives higher-curvature corrections
we are compelled to extremize a quantity which is no longer given by the area but a higher-derivative functional.
Hence, in order to find the extremal surface that yields the correct value of the entanglement entropy,
we must include an additional boundary condition to the problem.
We claim that the additional condition can be fixed by demanding that the
relationship between the bulk depth and the size of the entangling region
is the one induced by geodesics, we call this the \textit{free-kick condition}.
We implement this prescription in the computation of the entanglement entropy of the hairy black hole in new massive gravity,
and find a perfect agreement with conformal field theory expectations.
}
\maketitle
\onehalfspace

%%%%%%%%%%%%%%%%%%%%%%%%%%%%%%%%%%%%%%%
%%%%%%%%%%%%%%%%%%%%%%%%%%%%%%%%%%%%%%%
%%%%%%%%%INTRODUCTION%%%%%%%%%%%%%%%%%%%
%%%%%%%%%%%%%%%%%%%%%%%%%%%%%%%%%%%%%%%
\renewcommand{\thefootnote}{\arabic{footnote}}
\setcounter{footnote}{0}
\section{Introduction}
\label{intro}

A major force driving much progress in high-energy theoretical physics is the AdS/CFT correspondence or gauge/gravity duality \cite{Maldacena:1997re}. 
This duality relates systems that are different in many respects by providing
an accurate dictionary to relate quantities and phenomena in one system to those in the other.
More precisely, with the help of the correspondence, we can formulate questions
regarding quantum gravity in asymptotically AdS spacetimes as problems in a lower-dimensional
gauge theory, and vice versa. Moreover, \textit{hard} problems in one side of the duality are
translated into \textit{simple} problems in the other. This feature of the duality allows us to
treat analytically questions that were hitherto out of the grasp of our tools, e.g. strongly coupled field theories.
At the same time, this aspect of the correspondence makes it hard to prove its validity in general.
In spite of that, the AdS/CFT correspondence has been subjected to numerous highly nontrivial tests and it has passed them with flying colors.  

Entanglement entropy is a very interesting quantity that can be defined for any quantum system.
Roughly speaking, the entanglement entropy of a subsystem quantifies the amount of information that we would forfeit
if we were to lose access to the rest of the system. It must be said, that analytic computations of entanglement entropy,
in general, can be rather hard. However, there is a great body of literature with many interesting results, see, e.g., \cite{Casini:2009sr, Calabrese:2009qy}.
In recent years, there has been a renewed interest in the study of entanglement entropy.
This is partly due to the reformulation of the problem, under the light of the AdS/CFT correspondence, by Ryu and Takayanagi (RT) \cite{Ryu}.
Their proposal has been applied with great success to a wide variety of systems.
Moreover, it has been generalized in many different respects;
of particular interest for this work has been the extension of this prescription
to theories whose gravitational dual receives higher curvature contributions \cite{Solodukhin:2008dh, Bhattacharyya:2013gra, Camps:2013zua, Dong:2013qoa, Fursaev:2013fta}.

Three-dimensional gravity has played an important role in the holographic study of entanglement entropy.
This is no doubt related to the great deal of analytic understanding that we possess of both 3d gravity and 2d conformal field theory (CFT).
For example, the entanglement entropy for an interval in (1+1)-dimensional CFTs has been computed exactly both at zero \cite{Holzhey:1994we}
and at finite temperature \cite{Calabrese}, these two results have been matched using the RT prescription.
Three-dimensional higher-derivative gravities have been the subject of ample interest  in recent years. 
In this work we focus on one such theory, namely, new massive gravity (NMG) \cite{Bergshoeff:2009hq}.
It is a well-known fact that gravitons in three dimensions carry no degrees of freedom; however,
if we endow the graviton with a mass it will carry two degrees of freedom instead.
Thus, we expect a theory such as NMG to have attractive new properties to study.
Indeed, this theory admits interesting nontrivial solutions and, specifically,
black holes with an extra {\it gravitational hair} parameter \cite{Oliva:2009ip}.
Some properties of these black holes have been studied in the
context of holographic dualities \cite{Bergshoeff:2009aq, Giribet:2009qz, Giribet:2010ed, Kwon:2011jz, deBuyl:2013ega, Fareghbal:2014kfa}.
In the present work, we wish to further these studies by investigating in detail the
holographic entanglement entropy of an asymptotically AdS hairy black hole in NMG.

Due to the additional hair parameter of these black holes we encounter a puzzle during the calculation of holographic entanglement entropy.
Indeed, the hair parameter leads to an ambiguity in finding the extremal surface used for computing the entanglement entropy holographically,
since the geodesic is not the entangling curve anymore.
The question then is what is the appropriate boundary conditions to find the entangling curve?
This is the main motivation of this paper.
We make a proposal for this issue and test it successfully against CFT expectations.
We claim that the ambiguity can be fixed by demanding that the
relationship between the bulk depth and the size of the entangling region
is the one induced by geodesics, we call this the \textit{free-kick condition}.

The plan of this paper is as follows:
In Sec.\,\ref{RevMG} we review massive gravity in (2+1)-dimensions and then we restrict ourselves to NMG theory.
We present the asymptotically AdS hairy black hole solution of this theory along its thermodynamical properties.
In Sec.\,\ref{HEENMG} we mention the known facts about entanglement entropy in (1+1)-dimensional CFT and
its geometrical interpretation in the AdS/CFT context.
We then introduce the entropy functional used for computing the entanglement entropy
in a general higher-derivative gravity theory. We also fix our notations once and for all.
In Sec.\,\ref{GeoOTT} we study the geodesics in an asymptotically AdS hairy black hole background and derive an analytic expression for them.
In Sec.\,\ref{HEEBTZ} we use these results and calculate the entanglement entropy of the BTZ black hole in NMG.
In Sec.\,\ref{HEEOTT} we perform an exact analysis of holographic entanglement entropy for the asymptotically AdS hairy black hole solution of NMG.
We state our proposal to impose constraint on the extremal surface used for computing entanglement entropy holographically.
We then proceed to interpret our results for holographic entanglement entropy from the boundary field theory perspective and verify it.
We also comment on the closed extremal surfaces in this background.
In Sec.\,\ref{Conclusions} we summarize our results and discuss the possible future directions.
In Appendix \ref{Elliptic app}, we briefly introduce the special functions that we use in the rest of the paper.

\section{Review of new massive gravity}
\label{RevMG}
In this work we wish to study the entanglement entropy of a (2+1)-dimensional asymptotically AdS hairy black hole.
This black hole is a solution of a theory called NMG, which we discuss briefly in the present section. 
Consider the three-dimensional massive gravity described by the action
\be\label{massiveAc}
S=\frac{1}{16\pi G}\int d^{3}x \sqrt{-g}\left[R-2\Lambda
+\frac{1}{2\mu_{\rm cs}}{\rm CS}(\Gamma)
+\frac{1}{m^{2}}K\right],
\ee
where
\begin{align}
{\rm CS}(\Gamma)&=\varepsilon ^{\alpha \beta \gamma}\Gamma _{\;\alpha \sigma }^{\rho }
\left(\partial _{\beta }\Gamma _{\;\gamma \rho}^{\sigma }
+\frac{2}{3}\Gamma _{\;\beta \eta }^{\sigma }\Gamma _{\;\gamma \rho}^{\eta }\right)\, ,\\
K&=R_{\mu\nu}R^{\mu\nu}-\frac{3}{8}R^{2}\nonumber\, .
\end{align}
The first term, ${\rm CS}(\Gamma)$, is the gravitational Chern-Simons term of topological massive gravity,
while $K$ is the higher curvature term introduced in \cite{Bergshoeff:2009hq} which defines the NMG theory.
The field equations for \eqref{massiveAc} are 
\be\label{fieldeq}
R_{\mu \nu }-\frac{1}{2}Rg_{\mu \nu }
+\Lambda g_{\mu \nu }+\frac{1}{2m^{2}}
K_{\mu \nu }+\frac{1}{\mu_{\rm cs}}C_{\mu \nu}=0\, ,
\ee
where $C_{\mu \nu}$ is the Cotton tensor given by
\be
C_{\mu \nu}=\varepsilon _{\mu }^{~\alpha \beta }\nabla_{\alpha }
\left(R_{\beta \nu }-\frac{1}{4}g_{\beta\nu}R\right)\, ,
\ee
and $K_{{\mu \nu }}$ reads
\begin{equation}
K_{\mu \nu }=2\square {R}_{\mu \nu }-\frac{1}{2}\nabla _{\mu }\nabla _{\nu }{R}
-\frac{1}{2}\square {R}g_{\mu \nu }+4R_{\mu \alpha \nu \beta }R^{\alpha\beta }
-\frac{3}{2}RR_{\mu \nu }-R_{\alpha \beta }R^{\alpha \beta }g_{\mu\nu }
+\frac{3}{8}R^{2}g_{\mu \nu }\, ,
\end{equation}
and fulfills $K=g^{\mu \nu }K_{\mu\nu }$.

In the rest of the paper we will consider the parity preserving action, i.e. $\mu_{\rm cs} \to \infty$.
In the following section we will take the value of $m$ such that the NMG admits an asymptotically AdS hairy black hole.

\subsection{Asymptotically AdS hairy black hole at $m^2L^2=1/2$} \label{OTT sec}

 For the special point in the parameter space $m^{2}L^{2}=1/2$,
$L$ being the AdS radius, the above theory \eqref{massiveAc} admits the following black hole solution
which obeys weakened AdS$_3$ asymptotic boundary conditions \cite{Oliva:2009ip},
\be\label{metric}
ds^2=-f(r) dt^2+f(r)^{-1} dr^2+r^2d\phi^2\, ,
\ee
with
\be\label{warp}
  f(r)=- \mu+b r+\frac{r^2}{L^2}\, .
\ee
Notice that the above black hole solution has vanishing Cotton tensor
and hence it satisfies the full massive gravity equations of motion (\ref{fieldeq}).
This black hole is labeled by two parameters: the integration constant $\mu$ and the gravitational hair parameter $b$.
The mass of the black hole is given by
\be
M=\frac{1+\mu}{4 G}\, .
\ee
In the $b\to 0$ limit the solution is reduced to the BTZ black hole,
while in the case of nonvanishing $b$ the geometry develops a curvature singularity at $r=0$.
In the case of $b>0$ there is a single event horizon located at
\be\label{horizon}
r_{+}=\frac{1}{2}\left(-b L^2+\sqrt{b^2 L^4+4 \mu  L^2}\right)\, ,
\ee
 provided the bound $\mu\geq 0$ is satisfied.
We can associate a Hawking temperature and entropy to it
\begin{align}
T&=\frac{1 }{4\pi L}\sqrt{b^2 L^2+4\mu}\, ,\label{Temperature}\\
S&=\frac{\pi L}{2G}\sqrt{b^2 L^2+4\mu}\, .  \label{Entropy}
\end{align}
Notice that the entropy of the BTZ black hole, i.e., $b=0$, is twice the one obtained in pure Einstein gravity.
Henceforth, we refer to \eqref{metric} along with \eqref{warp} as the Oliva, Tempo, Troncoso (OTT)
black hole.

\section{Holographic entanglement entropy for NMG}
\label{HEENMG}

Entanglement is the characteristic trait of quantum mechanics \cite{Sch} and as such it is important to devise concrete methods of quantifying it.
For example, consider the following simple idea.
Imagine that the system of interest is in a pure state $|\Psi\rangle$ and suppose that we wish to quantify
the entanglement between a subsystem $A$ and its complement $\bar A$.
As a first step we could find the reduced density matrix obtained by tracing out the degrees of freedom in $\bar A$
\be
\rho_A=\Tr_{{\cal H}_{\bar A}}|\Psi\rangle\langle \Psi|\, .
\ee
Once we have constructed this matrix, we notice that if there is any entanglement between the
degrees of freedom in $A$ and those in $\bar A$ the system appears, to an observer having access only to $A$, to be in a mixed state.
If that is the case, then the Von Neumann entropy of $\rho_A$
\be
S_{{\rm EE}}(A)=-\Tr(\rho_A\log \rho_A)\, ,
\ee
is nonvanishing. We refer to this quantity as the entanglement entropy of $A$. 

In practice it is rather difficult to carry out the procedure outlined above due to technical limitations.
In spite of this, many general results have been found (e.g., strong subadditivity, area law divergence, etc.)
and a great deal of progress has been made. Of particular interest to us is the fact that in
a (1+1)-dimensional CFT the entanglement entropy of an interval of length $\ell$ can be computed analytically and it is given by
\be \label{CC1}
S_{{\rm EE}}(A)=\frac{c}{3}\log\left(\frac{\ell}{\epsilon}\right),
\ee
where $c$ is the central charge of the CFT and $\epsilon$ an ultraviolet cutoff \cite{Holzhey:1994we, Calabrese}.
Moreover, if we put the system at finite temperature and assume periodic boundary conditions, we find 
\be \label{CC2}
S_{{\rm EE}}(A)=\frac{c}{3}\log\left[\frac{\beta}{\pi\epsilon}\sinh\left(\frac{\pi \ell}{\beta}\right)\right],
\ee
where $\beta$ is the inverse temperature \cite{Calabrese}.

\begin{figure}
    \centering
    \includegraphics[scale=0.45]{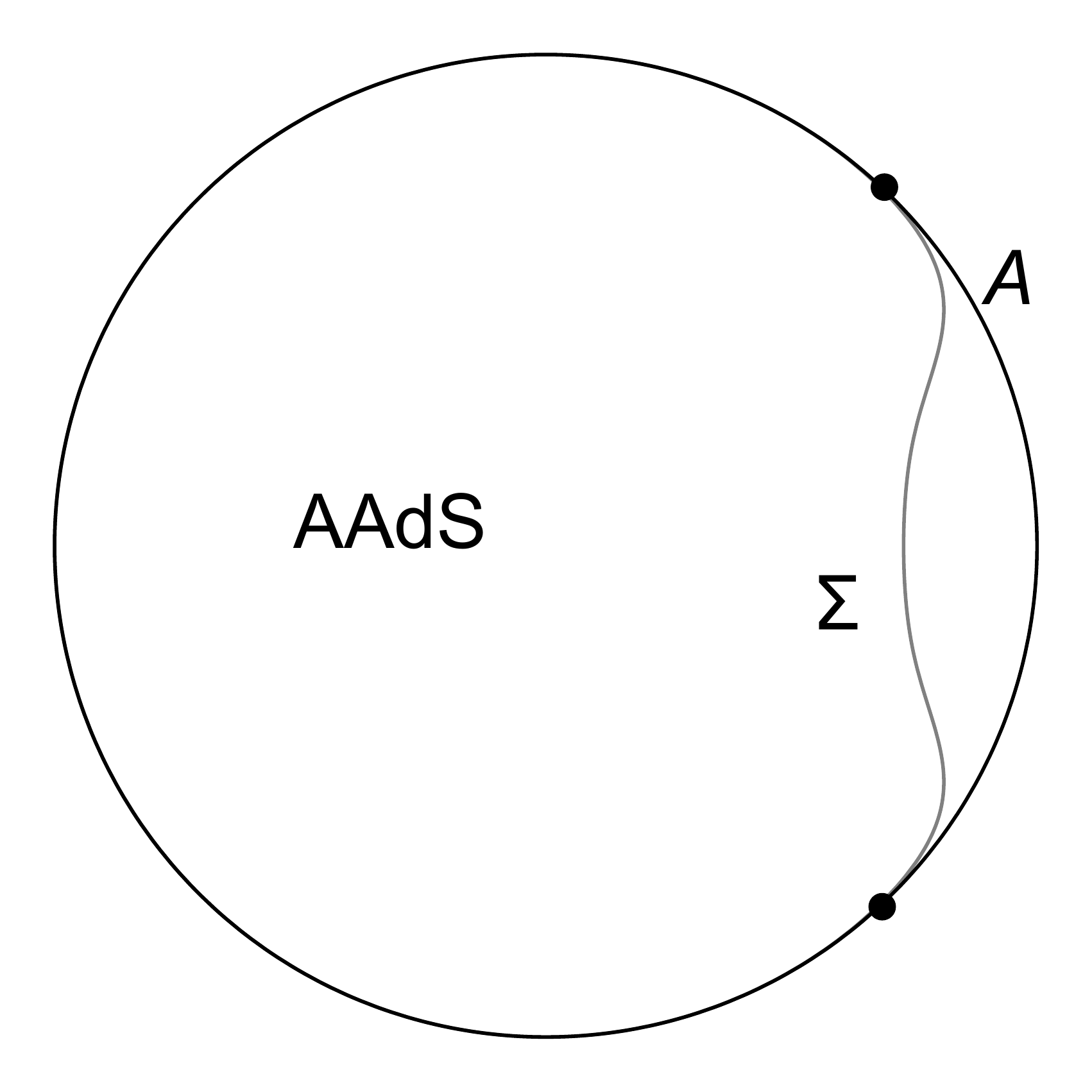}
\caption{The Ryu-Takayanagi prescription; $A$ is the entangling region, and $\Sigma$ is the extremal surface in the asymptotically AdS background.}\label{RTf}
\end{figure}

The study of entanglement entropy has been living a period of renaissance during the past few years,
mainly thanks to the holographic reformulation of the problem due to Ryu and Takayanagi \cite{Ryu}.
In its original form, the RT proposal claims that for a theory with an
Einstein gravity dual the computation of the entanglement entropy can
be recast as a Plateau problem in an asymptotically AdS spacetime. From a practical point of view,
in order to compute the entanglement entropy for a subsystem $A$ in the boundary theory one needs to extremize the functional
\be\label{RT1}
S_{{\rm EE}}(A)=\frac{1}{4G}
\int_{\Sigma} d^{d-1}y\sqrt{h}\, ,
\ee
where $\Sigma$ is a codimension two hypersurface which we demand to be anchored
at $\partial A$ (see Fig.\;\ref{RTf}), and $h$ is the induced metric on $\Sigma$.
Hence the entanglement entropy of $A$  is given by 
\be\label{RT2}
S_{{\rm EE}}(A)=\frac{1}{4G}\min_{\partial A=\partial \Sigma} \text{Area}\left(\Sigma\right)\, .
\ee

Notwithstanding its great efficacy, the prescription (\ref{RT2}) has two important limitations.
First of all, it is valid for static spacetimes only, this issue has been addressed
by Hubeny, Rangamani and Takayanagi by developing a covariant formulation
of the RT proposal \cite{Hubeny:2007xt}; in this paper we focus on static solutions so
we will not discuss the details of this development. On the other hand, experience
with black hole entropy \cite{Wald:1993nt} teaches us that in
the presence of higher curvature corrections, entropy computations must be modified;
this general wisdom should obviously apply to the RT prescription.
This question has been discussed recently in
\cite{Bhattacharyya:2013gra, Camps:2013zua, Dong:2013qoa,Fursaev:2013fta, Solodukhin:2008dh}.
In the following we summarize the results in this direction that will be relevant for the forthcoming sections.

Let us consider a general four-derivative gravity action
\be\label{four der}
S=\frac{1}{16\pi G}\int d^{d+1}x \sqrt{-g}\left[R-2\Lambda+c_1 R^2+c_2 R_{\mu\nu}R^{\mu\nu}+c_3 R_{\mu\nu\rho\sigma}R^{\mu\nu\rho\sigma}\right].
\ee
For these models, the entanglement entropy is still determined using a codimension
two surface extending into the bulk as in Fig.\;\ref{RTf}.
However, the shape that $\Sigma$  is compelled to take in order to yield the correct
value of the entanglement entropy is now determined by extremizing the functional \cite{Fursaev:2013fta, Camps:2013zua,Dong:2013qoa}
\be\label{EE HC}
S_{{\rm EE}}=\frac{1}{4G}
\int_{\Sigma} d^{d-1}y\sqrt{h}
\left[1+2 c_1 R +c_2\left(R_{||}-\frac{1}{2}{\cal K}^{2}\right)+2 c_3\left(R_{||\,||}-\Tr\left({\cal K}\right)^{2}\right)
\right]\, ,
\ee
instead of (\ref{RT1}).
In the following we explain how to obtain the different geometric quantities entering into this functional.

The first step is to find a basis for the vector space normal to the surface $\Sigma$.
In particular,  we choose basis vectors $n_{(0)}^\mu$ and $n_{(1)}^\mu$ such that 
\be 
g_{\mu\nu}n_{(\alpha)}^\mu n_{(\beta)}^\nu=\eta_{\alpha\beta}\, ,
\ee
where $\eta_{\alpha\beta}$ is the two-dimensional Minkowski metric.
Once we have found these vectors it is possible to construct all the ingredients entering Eq.\,\eqref{EE HC}.
The induced metric is simply
\be 
h_{\mu\nu}= g_{\mu\nu}-\eta^{\alpha\beta}\left(n_{(\alpha)}\right)_\mu\left(n_{(\beta)}\right)_\nu\, ,
\ee
while the relevant contributions from the ambient Riemann curvature read
\be
 R_{||\,||}=\eta^{\alpha\delta}\eta^{\beta\gamma}   \left(n_{(\alpha)}\right)^\mu\left(n_{(\delta)}\right)^\nu \left(n_{(\beta)}\right)^\rho\left(n_{(\gamma)}\right)^\sigma R_{\mu\nu\rho\sigma}\, ,
\ee
and
\be
 R_{||}=\eta^{\alpha\beta}\left(n_{(\alpha)}\right)^\mu\left(n_{(\beta)}\right)^\nu R_{\mu\nu}\, .
\ee
In turn, the extrinsic curvature is given by
\be
\left({\cal K}_{(\alpha)}\right)_{\mu\nu}=h_{\mu}^{\;\lambda}h_{\nu}^{\;\rho}\nabla_\rho \left(n_{(\alpha)}\right)_\lambda\, , 
\ee
where $\nabla$ is the covariant derivative \textit{wrt} $g_{\mu\nu}$.
The contractions of ${\cal K}_{(\alpha)}$ entering the functional  (\ref{EE HC}) are
\be\label{K sq}
{\cal K}^2\equiv \eta^{\alpha\beta}( {\cal K}_{(\alpha)})_{\mu}^{\;\mu}( {\cal K}_{(\beta)})_{\nu}^{\;\nu}\, ,
\ee
and
\be
\Tr\left({\cal K}\right)^2\equiv \eta^{\alpha\beta}( {\cal K}_{(\alpha)})_{\mu}^{\;\nu}( {\cal K}_{(\beta)})_{\nu}^{\;\mu}.
\ee

In the following we shall be concerned with solutions to the NMG theory (\ref{massiveAc}).
In the notation of Eq.\,\eqref{four der} this theory corresponds to the coefficients
\be 
c_1=-\frac{3}{8m^2}\, , \qquad c_2=\frac{1}{m^2}\, ,\qquad c_3=0\, .
\ee
Therefore the functional (\ref{EE HC}) reduces to
\be\label{EE}
S_{{\rm EE}}=\frac{1}{4G}
\int_\Sigma d\tau\sqrt{h}
\left[1+\frac{1}{m^2}\left(R_{||}-\frac{1}{2}{\cal K}^{2}-\frac{3}{4}R\right)
\right]\, .
\ee
The computation of entanglement entropy using this functional is the task upon which we embark in the following sections.

\section{Geodesics in the OTT geometry}\label{geodesic OTT}
\label{GeoOTT}

According to the RT prescription (\ref{RT2}), the entanglement entropy for an interval $A$ in a (1+1)-dimensional CFT
corresponds to the length of a geodesic extending into the bulk attached to the ends of $A$.
In this paper we are interested in computing the entanglement entropy for a system dual to
the OTT background (\ref{warp}); as seen in Sec.\,\ref{OTT sec}, this background is a solution of a higher curvature theory, NMG.
Therefore, we do not expect the entanglement entropy and the geodesic length to match,
instead we expect (\ref{EE}) to yield the correct answer.
However, the geodesics in the OTT background will play an indirect yet important role
in the computation of  entanglement. Therefore, we devote the present section to the study of these curves. 

Geodesics are found by extremizing the functional 
\be\label{length}
I(A)=
\int_{\Sigma_A} d^{d-1}y\sqrt{h}\, ,
\ee
where $A$ is a spacelike interval at asymptotic infinity and $\Sigma_A$ has its endpoints fixed at $\partial A$.
We could either parametrize these curves by the angular or by the radial AdS coordinates,
hereafter we refer to the former as the \textit{boundary parametrization}
and to the latter as the \textit{bulk parametrization}.
Choosing the bulk parametrization, we find that the induced metric is given by
\be
h=r^2\phi'(r)^2-f(r)^{-1}\, ,
\ee
where $f(r)$ is given by Eq.\,\eqref{warp}. Notice that the functional (\ref{length}) does not depend explicitly on $\phi$ but only on its derivatives. Setting $L=1$, we find 
\be
\phi'(r)=\frac{1}{r\sqrt{(r-r_+)(r+r_+ +b)(a r^2-1)}}\, ,
\ee
where $a$ is an integration constant and $r_+$ is the horizon of the black hole (\ref{horizon}). 
Therefore, the profile of a geodesic reaching down to a radius $r_*\geq r_+$ is given by 
\be
\phi(r)= \int_{r_*}^r d\tilde r\; \phi'(\tilde r)\, .
\ee
The integration constant $a$ can be determined by imposing the boundary condition 
\be
\frac{d r}{d\phi}\bigg|_{r_*} =0\, ,
\ee
which implies that $a=r_*^{-2}$.
 Hence, the bulk-parametrized geodesics read 
\be
\phi(r)=\int_{r_*}^r d\tilde r \frac{r_*}{\tilde r\sqrt{(\tilde r-r_+)(\tilde r+r_+ +b)( \tilde r^2-r_*^2)}}\, .
\ee
 Performing this integration explicitly, we find  
\be\label{hairy geodesic}
\phi(r)=\frac{2}{r_{+}}\frac{r_{*}F (z|\eta)+\left(r_{+}-r_{*}\right)\Pi (n;z|\eta)}
{\sqrt{(r_{+}+r_{*}) (b+r_{+}+r_{*})}}\, ,
\ee
where $F (z |\eta)$ and $\Pi (n;z|\eta)$ are incomplete elliptic integrals of the first and the third kind, respectively and
\begin{subequations}
\begin{align}
n&=\frac{2 r_{+}}{r_{+}+r_{*}}\, ,\\
z&=\arcsin\left(\sqrt{\frac{(r-r_{*}) (r_{+}+r_{*})}{2 r_{*} (r-r_{+})}}\right)\,\label{z} ,\\
\eta&=\frac{2 r_{*} (b+2 r_{+})}{(r_{+}+r_{*}) (b+r_{+}+r_{*})}\, .
\end{align}
\end{subequations}
More details about these integrals can be found in Appendix \ref{Elliptic app}.

\begin{figure}
    \centering
    \includegraphics[scale=0.45]{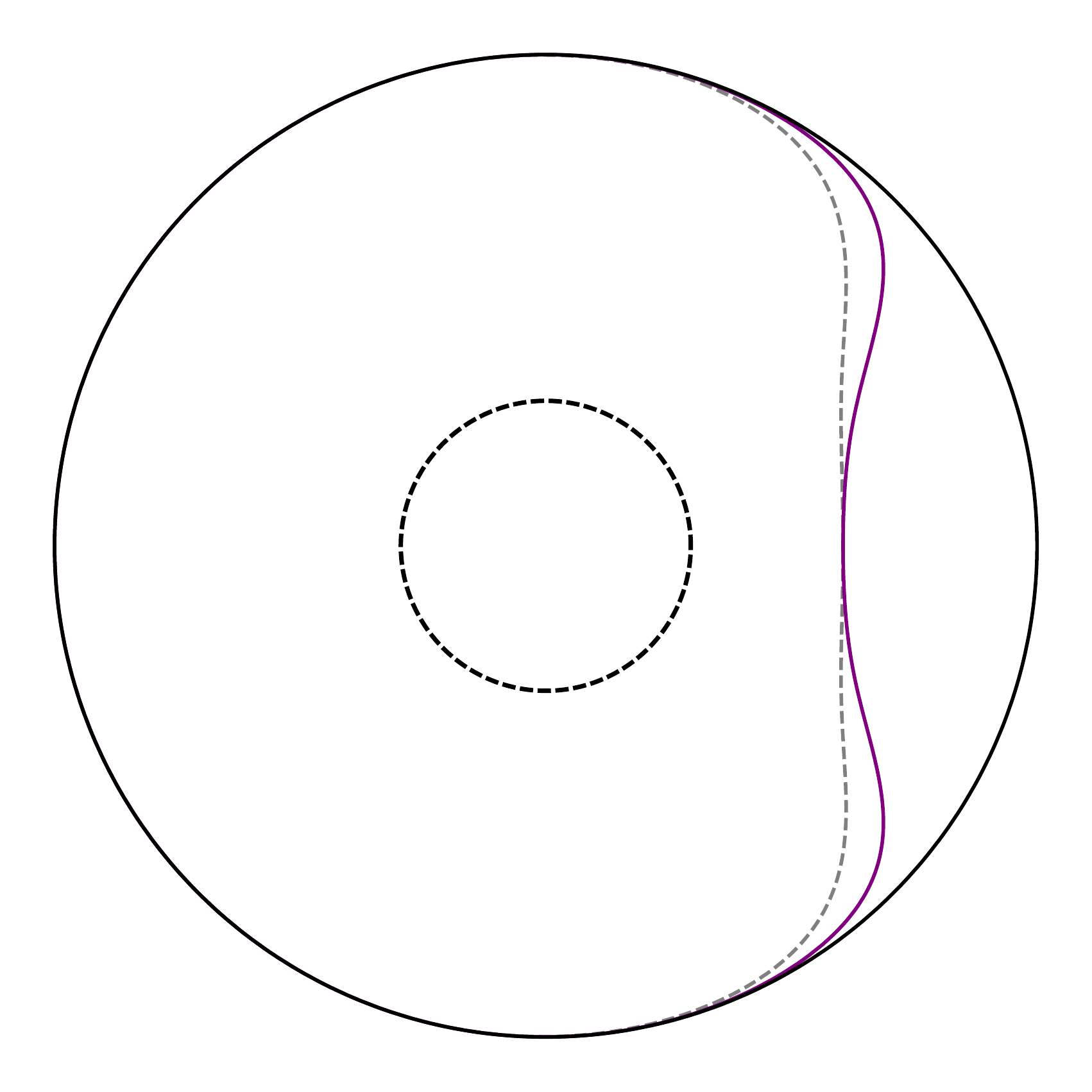}
\caption{The comparison between geodesics in the BTZ (dashed) and OTT (solid purple) geometries;
the radial coordinate is compactified using the map $r\to\arctan(r)$.
The outer (solid black) circle represents asymptotic infinity, and the inner (dashed) circle the horizon radius.}\label{OTT-BTZ}
\end{figure}

\begin{figure}
    \centering
    \includegraphics[scale=0.45]{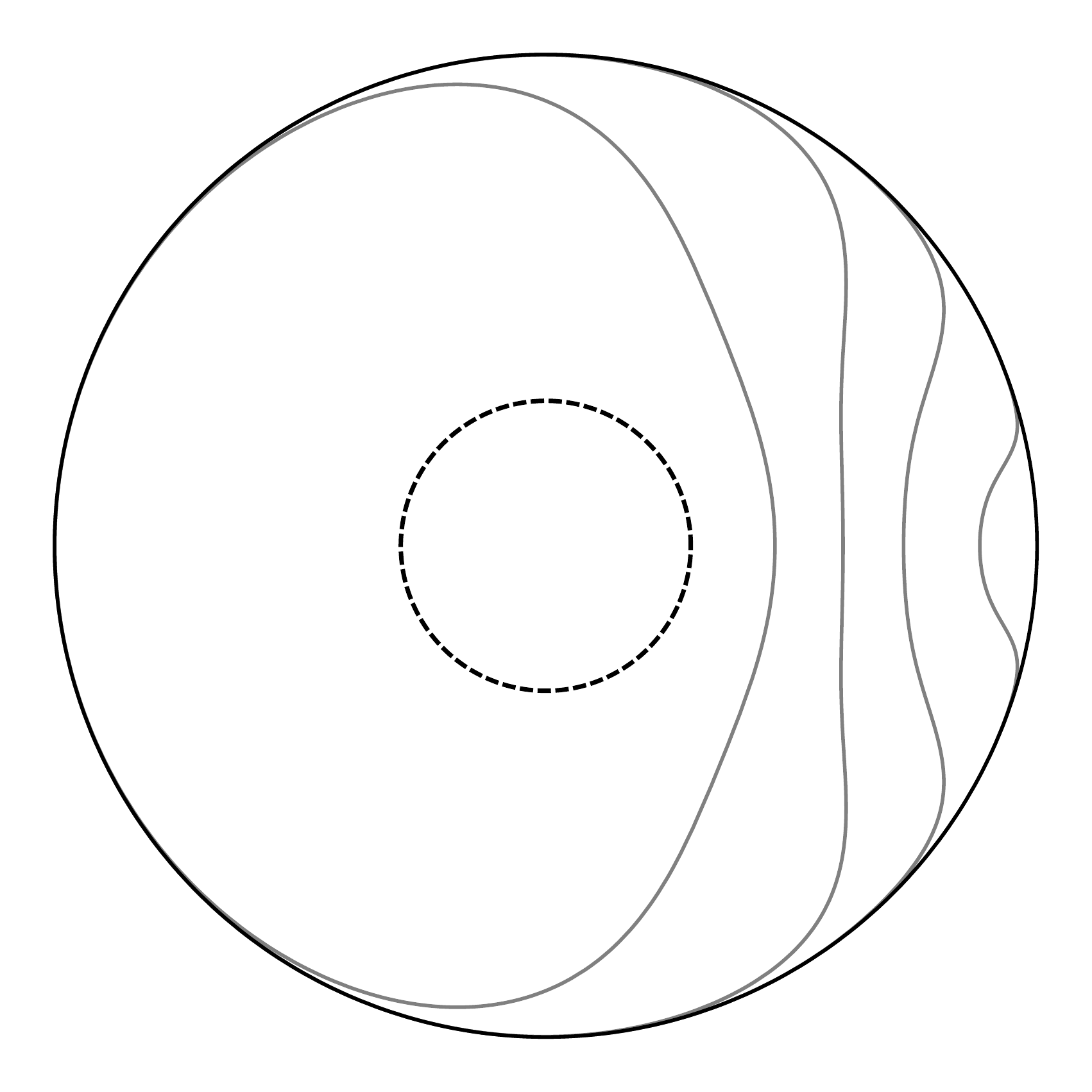}
\caption{Conformal diagram of entangling curves for different depths in the BTZ geometry.}\label{BTZ}
\end{figure}

Notice that in the bulk parametrization the depth $r_*$  reached by the geodesic is the tunable parameter (see Fig.\,\ref{BTZ}),
from the boundary point of view; however,
one would wish to specify the size of the entangling region $\tilde\phi$ instead.
It is clear that these two quantities are related by  
\be\label{boundary bulk}
\tilde\phi_{\rm geodesic}(r_*)= \phi(r,r_*)\Big|_{r\to\infty}\, .
\ee
To simplify notation, we drop the subscript \lq\lq{}geodesic\rq\rq{} when no confusion arises,
and unless otherwise stated.
Regretfully, it is quite complicated to solve this equation in general;
however, it is possible to do so by setting $b=0$,
in this case the OTT solution becomes the BTZ black hole and (\ref{hairy geodesic}) reduces to
(see Fig.\,\ref{OTT-BTZ} to compare the OTT and BTZ geodesics)
\be\label{geobtzphi}
\phi(r)= \frac{1}{r_+}\,\text{arccosh}\left(\sqrt{\frac{r_*^2-\left(\frac{r_+r_*}{r}\right)^2}{r_*^2-r_+^2}}\right)\, .
\ee
Here, it is straightforward to go from one parametrization to the other and the boundary parametrization is simply 
\be\label{geobtz}
r(\phi )=r_+ \left(1-\frac{\cosh ^2\left(r_+ \phi \right)}{\cosh ^2\left(r_+\phi_0 \right)}\right)^{-\frac{1}{2}}\, ,
\ee
where 
\be\label{phi0btz}
\phi_0=\frac{1}{r_+}\,\text{arccosh}\left(\frac{r_*}{\sqrt{r_*^2-r_+^2}}\right)\, .
\ee
Using Eq.\,\eqref{boundary bulk} we observe that $\tilde\phi=\phi_0$.
If $b\neq 0$ this relationship is more complicated but the relevant information can be
extracted systematically, we will come back to this issue in Sec.\,\ref{HEEOTT}. 

\section{Entanglement entropy of the BTZ black hole in NMG}
\label{HEEBTZ}

Before attempting to compute the entanglement entropy of the OTT black hole,
we set $b=0$ and perform this computation for the BTZ black hole.
The first step is to find the geometric quantities entering Eq.\,\eqref{EE},
the induced metric $h$ and $R_{||}$ are given by
\begin{align}
h=\frac{1}{\frac{r^2}{L^2}-\mu}+r^2 \phi '(r)^2\, ,\qquad
R_{||}&=-\frac{4}{L^2}\, ,
\end{align}
while the contraction (\ref{K sq}) of the extrinsic curvature reads
\begin{align}
{\cal K}^{2}(r)&=\frac{\left[2 L^2 r \left(r^2-\mu  L^2\right) \phi ''(r)
+2 r^2 \left(r^2-\mu  L^2\right)^2 \phi '(r)^3+\left(6 L^2 r^2
-4 \mu  L^4\right) \phi '(r)\right]^2}
{4 L^2 \left[r^2 \big(r^2-\mu  L^2\right) \phi '(r)^2+L^2\big]^3}\, .
\end{align}
Plugging these into (\ref{EE}) and carrying out the variation we find a complicated equation of motion for $\phi(r)$.
Interestingly enough, the solutions of the geodesic equation \eqref{geobtzphi}
solve also these more convoluted equations of motion as pointed out in \cite{Erdmenger:2014tba}\footnote{For the BTZ geometry,
the higher-derivative terms contribution is topological and can be written as a total derivative.
Therefore, the holographic entanglement entropy
is still determined by an extremal length curve in the bulk.}.

Notice that although the relevant extremal curves are the same as in Einstein gravity it is not
their length that we must consider, instead we ought to insert \eqref{geobtzphi}
into \eqref{EE} and compute the integral. By doing so,
we find that the entanglement entropy of the BTZ black hole in NMG is given by, setting $L=1$,
\begin{align}
S_{\rm EE}&=\frac{1}{ G}
\int_{r_{*}}^{r_{\epsilon }}
\frac{ r}{\sqrt{\left(r^2-r_{+}^2\right)\left(r^2-r_{*}^2\right)}}\, dr\nonumber\\
&=\frac{1}{G}\log \left(\sqrt{r^2-r_{+}^2}+\sqrt{r^2-r_{*}^2}\right)\Bigg|_{r_{*}}^{r_{\epsilon }}\, ,
\end{align}
where we introduced an ultraviolet cutoff $r_\epsilon\gg 1$.
Finally, using \eqref{phi0btz}, we can rewrite this expression in the boundary parametrization. We replace $r_*$ using  
\be
r_{*}^2=\frac{r_{+}^2}{1-\cosh ^{-2}\left(r_+ \tilde\phi \right)}\, ,
\ee
and obtain
\be
S_{\rm EE}=\frac{1}{G}\log \left[\frac{2r_\epsilon}{r_+}\sinh \left(r_+\tilde\phi\right)\right]\, .
\ee
Finally, introducing the quantities
\be
\beta=\frac{2\pi}{r_{+}}\, ,\qquad
\epsilon=\frac{1}{ r_{\epsilon }}\, ,\qquad
\ell=2\tilde\phi\, ,
\ee
we find
\be
S_{\rm EE}=\frac{1}{G}\log \left[\frac{\beta}{\pi \epsilon}\sinh \left(\frac{\pi \ell}{\beta }\right)\right]\, .
\ee
If we compare this result with (\ref{CC2}) we find that the expressions match if 
\be\label{c charge}
c=\frac{3}{G}\, .
\ee
The reader might find this puzzling since this $c$ seems to be twice the value of
the central charge computed by Brown and Henneaux \cite{Brown:1986nw}.
However, this contradiction is only apparent since the central charge for NMG
 is given by \cite{Bergshoeff:2009aq,Giribet:2009qz}
\be\label{c NMG}
c=\frac{3}{2G}\left(1+\frac{1}{2m^2}\right)\, ,
\ee
which at the special point  $m^2=1/2$ is precisely \eqref{c charge}.
Given that one can always find a BTZ black hole solution 
in NMG regardless of the point in parameter space,
the computation above can be replicated for arbitrary $m$; doing so, we find 
\be
S_{\rm EE}=\frac{1}{2G}\left(1+\frac{1}{2m^2}\right)\log \left[\frac{\beta}{\pi \epsilon}\sinh \left(\frac{\pi \ell}{\beta }\right)\right]\, ,
\ee
as expected. It is pleasant to see how (\ref{EE}) captures this shift of the central charge in a nontrivial manner.

\section{Entanglement entropy of the OTT black hole}
\label{HEEOTT}
The time for computing the entanglement entropy of the OTT black hole has arrived.
This computation will exhibit all the intricacies of calculating entanglement entropy in
a higher curvature theory, yet it is simple enough so that analytic expressions can be found.
The entropy corresponding to closed entangling curves was computed in \cite{Erdmenger:2014tba}.
Below, we focus on the case where these curves are anchored at infinity.
We shall see that to deal with this case it is necessary to introduce a new prescription to determine the boundary conditions.
Once again, we kick off by computing the geometric quantities entering \eqref{EE}.
The induced metric $h$ and $R_{||}$ are given by
\begin{align}\label{OTT h}
h&=\frac{1}{r \left(b+\frac{r}{L^2}\right)-\mu}+r^2 \phi '(r)^2\, , \nonumber\\
R_{||}&=-\frac{b L^2}{2 r \big[r^2 \phi '(r)^2 \left(b L^2 r-\mu  L^2+r^2\right)+L^2\big]}
-\frac{b}{r}-\frac{4}{L^2}\, ,
\end{align}
while the extrinsic curvature contraction \eqref{K sq} reads
\begin{align}\label{OTT K}
{\cal K}^{2}(r)&
=\frac{1}{4 L^2 \big[r^2 \phi '(r)^2 \left(b L^2 r-\mu  L^2+r^2\right)+L^2\big]^3}
\Big[2 L^2 r \phi ''(r) \left(b L^2 r-\mu  L^2+r^2\right)\nonumber\\&
+2 r^2 \phi '(r)^3 \left(b L^2 r-\mu  L^2+r^2\right)^2
+\phi '(r) \left(5 b L^4 r-4 \mu  L^4+6 L^2 r^2\right)\Big]^2\, .
\end{align}
For the present case, the Euler-Lagrange equation we need to solve is given by
\be\label{eom}
\frac{d^{2}}{dr^{2}}\left(\frac{\delta {\cal L}}{\delta \phi\rq{}\rq{}}\right)
-\frac{d}{dr}\left(\frac{\delta {\cal L}}{\delta \phi\rq{}}\right)
+\frac{\delta {\cal L}}{\delta \phi}=0\, ,
\ee
where ${\cal L}$  follows from \eqref{EE}, \eqref{OTT h} and \eqref{OTT K}.
Moreover, since this {\it Lagrangian} does not depend explicitly on $\phi(r)$, we have
\be\label{EL}
\frac{d}{dr}\left(\frac{\delta {\cal L}}{\delta \phi\rq{}\rq{}}\right)
-\frac{\delta {\cal L}}{\delta \phi\rq{}}=k\, ,
\ee
where $k$ is an integration constant. Putting everything together,
the equation of motion for $\phi(r)$ is given by the following complicated expression,
\begin{align}\label{ODE}
&\frac{1}{4 \sqrt{\frac{1}{r \left(b+\frac{r}{L^2}\right)-\mu }+r^2 \phi '(r)^2} \big[r^2 \phi '(r)^2 \left(b L^2 r-\mu  L^2+r^2\right)+L^2\big]^3}\nonumber\\&
\times \Bigg\{-L^4 r^2 \phi'(r)^3 \Big[r^2 \left(41 b^2 L^4+80 b L^2 r+44 r^2\right)
-4 \mu  L^2 r \left(23 b L^2+20 r\right)+56 \mu ^2 L^4\Big]\nonumber\\&
+4 r^6 \phi'(r)^7 \left(\mu  L^2+r^2\right)
\left(b L^2 r-\mu  L^2+r^2\right)^3+4 r^4 \phi'(r)^5 \left(-b L^2 r+6 \mu  L^2+2 r^2\right)\nonumber\\&
\Big[L r \left(b L^2+r\right)-\mu  L^3\Big]^2+4 L^6 r \Big[2 r \phi^{(3)}(r)
\left(b L^2 r-\mu  L^2+r^2\right)\nonumber\\&
+\phi''(r) \left(7 b L^2 r-4 \mu  L^2+10 r^2\right)\Big]
+4 L^4 \phi'(r) \Big[-5 r^4 \phi''(r)^2 \left(b L^2 r-\mu  L^2+r^2\right)^2\nonumber\\&
+2 b L^4 r+4 \mu  L^4
+8 L^2 r^2\Big]+8 L^4 r^3 \phi'(r)^2 \left(b L^2 r-\mu  L^2+r^2\right) \Big[r \phi^{(3)}(r)
\left(b L^2 r-\mu  L^2+r^2\right)\nonumber\\&
+\phi''(r) \left(-4 b L^2 r+3 \mu  L^2-5 r^2\right)\Big]\Bigg\}=k\, .
\end{align}

In the following, we proceed to solve Eq.\,\eqref{ODE} to linear order in $b$. Therefore, we write
\begin{align}
\phi(r)&=\phi^{(0)}(r)+b\,\phi^{(1)}(r)+{\cal O}\left(b^2\right)\, ,\\
k&=k^{(0)}+b\,k^{(1)}+{\cal O}\left(b^2\right)\, ,
\end{align}
and use \eqref{horizon} to express $\mu$ in terms of the horizon of OTT black hole and $b$.
Then we expand \eqref{ODE} around $b=0$.
Obviously, the zeroth order contribution reduces to the NMG-BTZ black hole,
hence, $\phi^{(0)}(r)$ is given by the geodesic profile \eqref{geobtzphi}
and $k^{(0)}=2r_{*}$. Nontrivial deviations from the geodesic path
appear already to linear order in $b$, setting $L=1$, we find 
\begin{align}\label{linear}
\phi^{(1)}(r)=-\frac{1}{\sqrt{r^2-r_*^2}}&\Bigg[-\sqrt{\frac{r-r_+}{r+r_+}}\,\frac{k^{(1)} r_+^2
\left(r+r_+\right)-r_* \left(r_*^2+r r_+\right)}
{2 r_+^2 \left(r_+^2-r_*^2\right)}
+c_1\nonumber\\&
+c_2\log \left(\sqrt{r^2-r_+^2}+\sqrt{r^2-r_*^2}\right)\Bigg]
-\frac{1}{2r_{+}}\phi^{(0)}(r)+c_3\, .
\end{align}
Observe that \eqref{linear} has four unknown constants $c_{1}$, $c_{2}$, $c_{3}$, and $k^{(1)}$ that we ought to fix.
Clearly, this is a consequence of the fact that the equations of motion \eqref{eom} is a fourth order differential equation,
but the pressing question is: How are we going to fix these extra constants?

\begin{figure}
    \centering
    \includegraphics[scale=0.45]{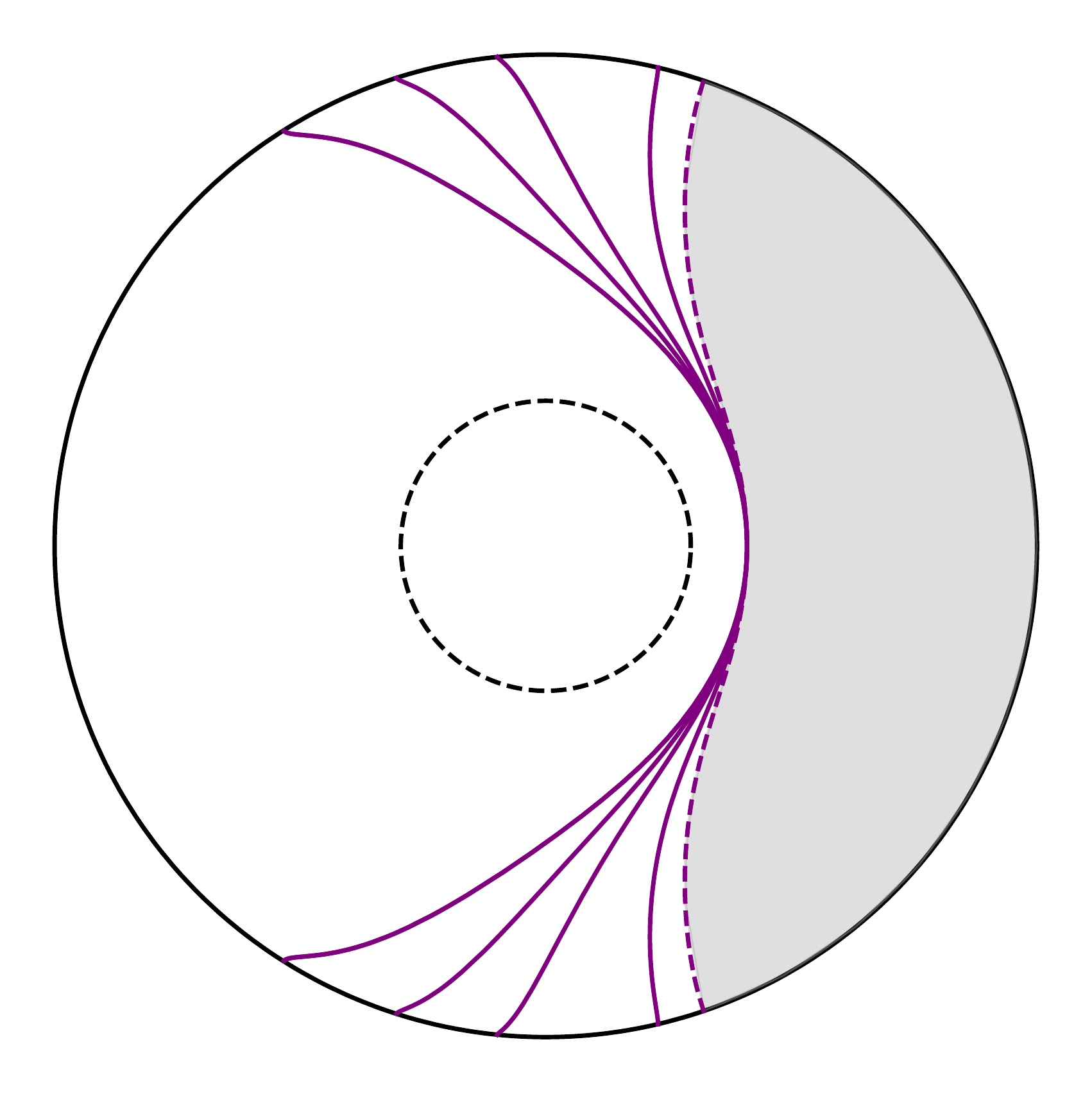}
\caption{Conformal diagram of extremal curves in the OTT geometry for different concavities (solid purple),
geodesic curve shadow (gray), and the free-kick extremal curve (dashed).}\label{OTT shadow}
\end{figure}

For a moment let us go back to the bulk parametrization.
Obviously we are also expected
to fix four independent constants there.
Three of them are rather easy to fix by setting
\begin{align}\label{conditions}
r(\phi)\bigg|_{\phi=0}=r_{*}\, ,\qquad
r'(\phi)\Big|_{\phi=0}=0\, ,\qquad
r'''(\phi)\Big|_{\phi=0}=0\, ,
\end{align}
we are already familiar with the first two conditions while the third one can be motivated by symmetry.
The missing parameter to fix is the \textit{initial acceleration} or the concavity of the entangling curve.
In Fig.\;\ref{OTT shadow} the different solid curves represent solutions complying with
\eqref{conditions} for different choices of $r''(0)$.
Hence, there seems to be an unfixed parameter in the game,
so the natural question now is whether there is any preferred initial acceleration.

It is at this point that the discussion in Sec.\,\ref{geodesic OTT} becomes relevant. We propose that 
this freedom can be fixed by demanding that the curve used for computing the
entanglement entropy holographically must not intersect the shadow
enclosed by the geodesic \eqref{hairy geodesic}
(this is the gray region in Fig.\;\ref{OTT shadow}).
More explicitly, we claim that the acceleration at $\phi=0$ has to be chosen such that  
\be\label{FK}
\tilde\phi(r_*)= \tilde\phi_{\rm geodesic}(r_*)\, .
\ee
This corresponds to the dashed line in Fig.\;\ref{OTT shadow}.
Hereafter, we refer to \eqref{FK} as the free-kick condition
since it is analogous to the problem that a football player has to solve when executing a free-kick.
Namely, given an object at rest at a fixed position, what is the acceleration necessary to hit a predetermined target.
In this analogy, the player is placed at the tip of the entangling curve,
while the target is given by the asymptotic value of the geodesic at spatial infinity.
Back in the boundary parametrization, the boundary conditions \eqref{conditions} read
\begin{subequations}\label{conditions 2}
\begin{align}
&\phi(r)\bigg|_{r=r_{*}}=0\, ,\\
&\frac{1}{\phi'(r)}\bigg|_{r=r_{*}}=0\, ,\\
&\frac{3 \phi''(r)^2-\phi'''(r) \phi '(r)}{\phi '(r)^5}\bigg|_{r=r_{*}}=0\, . 
\end{align}
\end{subequations}
From \eqref{linear} we see that the conditions \eqref{FK} and \eqref{conditions 2} are satisfied by setting
\begin{align}\label{const}
c_{1}&=\frac{r_{*}^2-k^{(1)} r_{+}^2}{2\,r_{+}^2 \sqrt{r_{*}^2-r_{+}^2}}\, ,\nonumber\\
c_{2}&=c_{3}=0\nonumber\, ,\\
k^{(1)}&=-\frac{3 r_*}{r_+}
-\frac{\sqrt{2 r_*} \left(r_+-r_*\right)^2}{3 r_+ \left(r_++r_*\right)^{3/2}}
~ F_1\left(\frac{3}{2};\frac{1}{2},\frac{3}{2};\frac{5}{2};
\sin(w),p\sin(w)\right)\nonumber\\
&+\frac{2 r_* \left[\left(r_+-r_*\right) r_*F(w|p)
+\left(r_++r_*\right)^2E(w|p)\right]}{r_+^2 \left(r_++r_*\right)}\, ,
\end{align}
where $E$ is the incomplete elliptic integral of the second kind \eqref{ei2},
$F_1$ is the Appell hypergeometric function \eqref{appel}, and 
\begin{align}
w=\arcsin\left(\sqrt{\frac{r_++r_*}{2r_*}}\right)\, ,\qquad
p=\frac{4 r_+ r_*}{\left(r_++r_*\right)^2}\, .
\end{align}
Additionally, we find that to linear order in $b$ the acceleration is given by
\be
r''(\phi)\Big|_{\phi=0}=-\frac{\phi ''(r)}{\phi '(r)^3}\bigg|_{r=r_{*}}=r_{*}\left(r_{*}^2-r_{+}^2\right)
+\frac{r_{*}}{2}\left(r_{*}k^{(1)}-2 r_{+}+r_{*}\right)b\, .
\ee

Finally, we insert \eqref{geobtzphi}, \eqref{linear}, and \eqref{const} into \eqref{EE}
and find the entanglement entropy of the OTT black hole to linear order in $b$
\begin{align}\label{OTT EE}
S_{\rm EE}(r_+,r_*)&=\frac{1}{2 G r_+^2}
\vast\{\frac{b \sqrt{r^2-r_+^2} \left[r^2 r_+^2 \left(r_+-k^{(1)} r_*\right)
+r \left(r_*^4-k^{(1)} r_+^3 r_*\right)+r_+ r_*^4-r_+^3 r_*^2\right]}
{r \left(r+r_+\right) \sqrt{r^2-r_*^2} \left(r_*^2-r_+^2\right)}\nonumber\\
&+\frac{b r_* \left(k^{(1)} r_+^2-r_*^2\right)}{\sqrt{r^2-r_*^2} \sqrt{r_*^2-r_+^2}}
+2 r_+^2 \log \left(\sqrt{r^2-r_+^2}+\sqrt{r^2-r_*^2}\right)
\vast\}\vast|^{r_{\epsilon}}_{r_*}+{\cal O}(b^2)\, . 
\end{align}
As can be seen from the above, the holographic entanglement entropy depends on the constant $k^{(1)}$,
at first order in $b$, which cannot be fixed unless extra boundary conditions are provided.
Notice that, for the BTZ geometry, $b=0$, the holographic entanglement entropy does not depend on $k^{(1)}$,
and therefore we do not need extra boundary conditions; this is consistent with the discussion in Sec.\,\ref{HEEBTZ}.
This expression seems to have a quite complicated dependence on the \textit{bulk parameters}
$r_+$ and $r_*$; we would like to know how \eqref{OTT EE} reads in
terms of the \textit{boundary parameters} $\beta$ and $\tilde\phi$. We test whether this expression matches the CFT expectation
\be\label{claim}
\frac{c}{3}\log\left[\frac{\beta}{\pi\epsilon}\sinh\left(\frac{\pi \ell}{\beta}\right)\right],
\ee
 with $\beta$ the inverse temperature of \eqref{Temperature} and $c$
the central charge \eqref{c charge} corresponds to the sought after $S_{\rm EE}(\beta, \ell)$.
We would like to emphasize that it happens since
we have fixed the constant $k^{(1)}$ using our proposed free-kick condition.
It is clear that other curves do not fit the CFT expression for
the entanglement entropy of an interval at finite temperature,
since the ordinary differential equation (ODE) is fourth order and we used four boundary
conditions to fix these freedoms for finding the entangling curve.
We proceed to verify this claim. First, we set $\ell=2\tilde\phi$ in $\eqref{claim}$ and use \eqref{FK}, i.e.,
\be
\tilde\phi(r_*)=\frac{2}{r_{+}}
\frac{r_{*}F (\tilde z|\eta)
+\left(r_{+}-r_{*}\right)\Pi (n;\tilde z|\eta)}
{\sqrt{(r_{+}+r_{*}) (b+r_{+}+r_{*})}}\, ,
\ee
where $\tilde z=\lim_{r\to \infty}z$, and $z$ is defined in \eqref{z}. Then we express $\beta$ in terms of the horizon $r_+$,
using \eqref{horizon} and \eqref{Temperature} we find
\be
\beta=\frac{4\pi}{b^2+2r_+}\, .
\ee
Inserting the above expressions into Eq.\,\eqref{claim} and expanding around $b=0$
we find that it matches \eqref{OTT EE} exactly at first order in $b$, this gives strong support to our claim.

Finally, we verify our claim numerically by adapting the algorithm discussed in
\cite{Bhattacharyya:2014oha} to the present case. In order to get useful numerical results,
we must compute the renormalized version of entanglement entropy
\cite{Casini:2004bw, Liu:2013una}, which for a (1+1)-dimensional CFT is given by 
\be\label{REE}
{\cal S}_{\rm EE}= \ell\frac{\partial}{\partial \ell} S_{\rm EE}\, .
\ee
The renormalized entanglement entropy corresponding to \eqref{claim} reads
\be\label{claim 2}
{\cal S}_{\rm EE}=\frac{c\, \ell}{3\beta}\coth\left(\frac{\pi \ell}{\beta}\right)\, .
\ee
Running a code that implements the free-kick condition alongside with the algorithm introduced in \cite{Bhattacharyya:2014oha}
we find that the ${\cal S}_{\rm EE}$ resulting from solving Eq.\,\eqref{ODE} matches \eqref{claim 2} within numerical error.

To finish this section, we briefly comment on the closed extremal surfaces in the OTT background. 
The black hole bifurcation surface $r_{+}$ is a closed extremal surface of the entanglement entropy functional \eqref{EE},
and the corresponding entanglement entropy is the Wald entropy of the black hole
\be
S_{\rm EE}^{+}=\frac{\pi L}{2G}\sqrt{b^2 L^2+4\mu}\, .
\ee
If we set $\phi'=\phi''=\phi'''=\phi''''=0$, we find another closed extremal surface given by
\be
r_a=L\sqrt{\mu }\, ,
\ee
whose corresponding EE is simply
\be
S_{\rm EE}^{a}=\frac{\pi  L \sqrt{\mu }}{G}\, .
\ee
Notice that  $r_{+}\geq r_{a}$.
In the absence of the gravitational hair parameter $b$,
i.e., the BTZ black hole, the bound is saturated and $S_{\rm EE}^{a}$ is exactly the Wald entropy of the BTZ black hole.

\section{Conclusions and outlook}
\label{Conclusions}

Since pure Einstein gravity in (2+1) dimensions admits few solutions,
one is compelled to consider more exotic actions like NMG.
The NMG theory admits a rich and interesting catalog of solutions such as black holes, wormholes, solitons, and kinks.
Hairy black holes are a characteristic of NMG and one would like to have a holographic interpretation for them.
The OTT black hole that we considered in this paper has the intriguing feature that 
geodesic curves do not correspond to the extremal surface used for computing the entanglement entropy holographically.
So, this work has established that if one would like to calculate the holographic entanglement entropy in higher-derivative gravity theories
one should first impose additional boundary conditions to find the extremal surface.
We demonstrated that this can be done by demanding that the entangling curve
must not intersect the region enclosed by the geodesic.

We then computed the holographic entanglement entropy of the OTT black hole.
We confirmed that the extremal surface which is fixed by proposing our boundary conditions
yields the known result for entanglement entropy in the (1+1)-dimensional CFT at finite temperature.

In the recent paper \cite{Alishahiha:2013zta}, the authors considered AdS wave solutions of NMG
which admit logarithmic modes in their Fefferman-Graham expansion.
These solutions are believed to be dual to logarithmic CFTs.
They calculated the holographic entanglement entropy for these backgrounds and
showed that the entanglement entropy has a new divergent term.
Beside the aforementioned, the NMG theory at the chiral point, where central charges vanish,
admits a logarithmic deformation of extremal BTZ as an exact solution.
The entanglement entropy for a CFT dual to the extremal BTZ black hole was computed recently \cite{Caputa:2013lfa}.
It is an interesting future problem to study the implications of the free-kick boundary conditions
to address the logarithmic deformation of extremal BTZ in boundary conformal field theory.

In this paper, to avoid even more cumbersome computations we focused on the static case; however,
it would be plausible to extend our current study to the rotating hairy black holes.
Due to the gravitational hair parameter the asymptotically AdS rotating hairy black holes have two different kind of extremal limits.
In one of them, both the temperature and the entropy vanish and, therefore, it defines a Nernst solution. It will be interesting to investigate the behavior of holographic entanglement entropy for these geometries.

Lastly, we would like to comment on some recent work which showed that one can put constraint on the extremal surfaces by imposing causality \cite{Erdmenger:2014tba}.
We proposed new boundary conditions to completely fix the extremal surface
used for computing holographic entanglement entropy in higher-derivative gravity theories.
One of the intriguing future directions is to try to understand the relation between the argument based on CFT causality \cite{Headrick:2014cta}
and our work. These are the avenues that we will try to explore in the near future.

\appendix

\section{Elliptic integrals}
\label{Elliptic app}

In this section we will shortly review the definitions of the special functions which we used in this paper.
The incomplete elliptic integrals can be written as
\begin{subequations}
\begin{align}
F(z|\eta)&=\int_0^z \frac{dt}{\sqrt{1-\eta \sin ^2(t)}}\, ,\\
E(z|\eta)&=\int_0^z \sqrt{1-\eta \sin ^2(t)} \, dt\,\label{ei2} ,\\ 
\Pi (n;z|\eta)&=\int_0^z \frac{dt}{\sqrt{1-\eta \sin ^2(t)} \left[1-n \sin ^2(t)\right]}\, .
\end{align}
\end{subequations}
The parameter $\eta$ is called the {\it modulus} of the elliptic integral and $z$ is the {\it amplitude angle}.
They range from $0\leq z\leq \frac{\pi}{2}$ and $0\leq \eta\le1$.

The Appel hypergeometric function $F_1$ is defined by the double series
\be\label{appel}
F_1\left(a;b_1,b_2;c;z_1,z_2\right)=
\sum _{k=0}^{\infty } \sum _{l=0}^{\infty }
\frac{z_1^k z_2^l \left(b_1\right)_k \left(b_2\right)_l (a)_{k+l}}{k! l! (c)_{k+l}}
\, ;\qquad\left| z_1\right| <1 \land \left| z_2\right| <1\, ,
\ee
where the Pochhammer symbol $(q)_n$ denotes the rising factorial,
\be
(q)_n=q(q+1)\dots (q+n-1)\, .
\ee

\acknowledgments
We are grateful to Shajid Haque for commenting on a draft of this paper.
We would  like to thank Mohsen Alishahiha, Arpan Bhattacharyya, Pawel Caputa,
Amin Faraji Astaneh, Mario Flory, Kevin Goldstein, Charlotte Sleight, and Amir Hadi Ziaie for illuminating conversations.
The research of A.V.O. is supported by the University Research Council of the University of the Witwatersrand.
The work of S.M.H. is supported in part by INFN. S.M.H. and A.V.O. wish to thank the hospitality of the ICTP
in Trieste where part of this research was conducted.
A.V.O. is grateful to the theory group at Universit\`a di Milano-Bicocca
for kindly hosting him during the concluding stages of this work.

\end{document}